\documentclass[12pt]{article}
\usepackage{amssymb,amsmath}
\usepackage{graphicx}
\usepackage{enumerate}
\usepackage{subfigure}
\usepackage{here}
\usepackage{hyperref}
\newcommand{\Ncal}{\mathcal{N}}
\newcommand{\del}{\partial}
\DeclareMathOperator*{\Tr}{{\rm Tr}}

\newcommand{\psib}{\bar{\psi}}
\newcommand{\thetab}{\bar{\theta}}
\newcommand{\phib}{\bar{\phi}}
\newcommand{\epsilonb}{\bar{\epsilon}}
\newcommand{\xib}{\bar{\xi}}
\newcommand{\Qb}{\overline{Q}}
\newcommand{\Db}{\overline{D}}
\newcommand{\deldel}[2]{\frac{\del #1}{\del #2}}
\newcommand{\im}{\mathrm{Im}}

\numberwithin{equation}{section}

\begin{document}

%%%%%%%%%%%%%%%%%%%%%%%%%%%%%%%%%%%%%%%%%%%%
\thispagestyle{empty}
\begin{flushright}
%KEK-TH 1468
%KUNS-2351\\
CALT 68-2916\\
OU-HET 780\\

%\\
%Aug 23, 2012
%arXiv:yymm.nnnn
\end{flushright}
\vskip1cm
\begin{center}
{\Large Supersymmetric boundary conditions \\ in \\ three-dimensional $\mathcal{N}=2$ theories}

%\bigskip\bigskip
%{\it  California Institute of
% Technology, Pasadena, California 91125, USA \\ Department of Physics Graduate School of Science, Osaka
% University, Toyonaka, Osaka 560-0043, Japan}
%\end{center}

\vskip2cm
Tadashi Okazaki$^{a,b}$\footnote{tadashi[at]theory.caltech.edu} and
 Satoshi Yamaguchi$^a$\footnote{yamaguch[at]het.phys.sci.osaka-u.ac.jp}

\bigskip\bigskip
$^a$
{\it Department of Physics, Graduate School of Science, Osaka University, Toyonaka, Osaka 560-0043, Japan}
\\
\vskip2mm
$^b$
{\it California Institute of Technology, Pasadena, California 91125, USA}
\end{center}

%%%%%%%%%%%%%%%%%%%%%%%%%%%%%%%%%%%%%%%%%%%%
\vskip2cm
\begin{abstract}
We study supersymmetric boundary conditions in three-dimensional
 $\mathcal{N}=2$ Landau-Ginzburg models and 
Abelian gauge theories. 
In the Landau-Ginzburg model the boundary conditions that 
preserve $(1,1)$ supersymmetry (A-type)  and $(2,0)$ supersymmetry (B-type)
on the boundary are classified in terms of subspaces of the target space (``brane'').
An A-type brane is a Lagrangian submanifold on which the imaginary part of the superpotential is constant,  while a B-type brane is a holomorphic submanifold 
on which the superpotential is constant. 
We also consider the $\mathcal{N}=2$ Maxwell theory with boundary and the Abelian duality.  
Finally we make some comments on $\mathcal{N}=2$ SQED with boundary condition and the mirror symmetry.
\end{abstract}

%%%%%%%%%%%%%%%%%%%%%%%%%%%%%%%%%%%%%%%%%%

\newpage
\setcounter{tocdepth}{2}
\tableofcontents
%%%%%%%%%%%%%%%%%%%%%%%%%%

\section{Introduction}
Quantum field theories with boundaries are worth investigating since they often play 
important roles in various fields of physics.  One of the most
interesting ones is a boundary CFT description of D-branes.  In
particular supersymmetric boundaries of two-dimensional $\Ncal=(2,2)$
quantum field theories have been well studied because they are good probes in the study of mirror symmetry \cite{Ooguri:1996ck,Hori:2000ck}.
More recently supersymmetric boundary conditions in four-dimensional $\mathcal{N}=4$
super Yang-Mills theories and their relation to S-duality have been examined by 
 \cite{Gaiotto:2008sa, Gaiotto:2008sd, Gaiotto:2008ak}.

It is also interesting to study supersymmetric boundary conditions
 in  three-dimensional supersymmetric field theories. One of the most
 attractive  motivations is that it will provide a description of M5-branes in terms
 of the boundary condition of M2-brane theories
 \cite{Townsend:1995af,Chu:1997iw,Berman:2009kj,Berman:2009xd}.  It will
 also be a useful tool to investigate various dualities in three dimensions such as
 mirror symmetry \cite{Intriligator:1996ex,Aharony:1997bx, deBoer:1997kr,Dorey:1999rb, Aganagic:2001uw} 
 and 3d-3d correspondence \cite{Terashima:2011qi,Terashima:2011xe,Dimofte:2011ju, Dimofte:2011py}.

\begin{figure}
\begin{center}
\includegraphics[width=8cm]{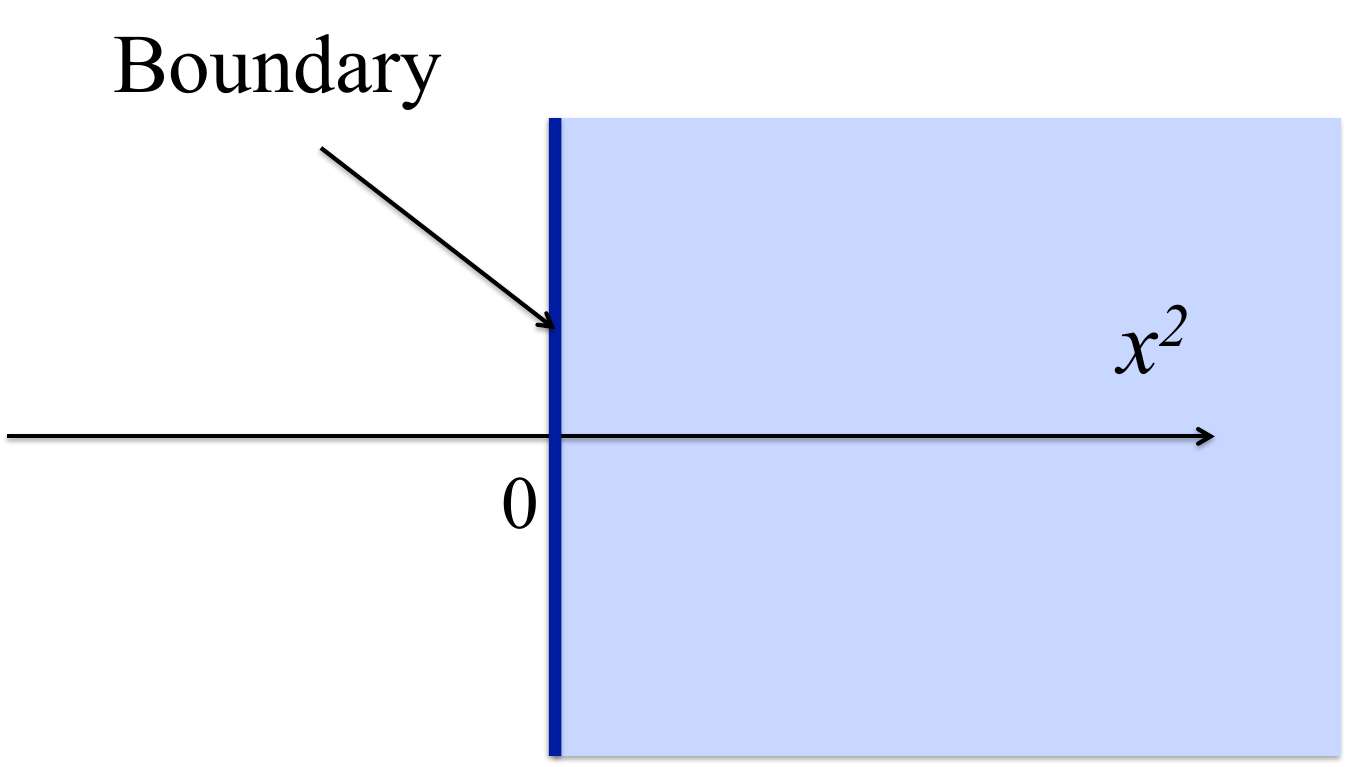}
\caption{The theory defined in $x^{2}\ge 0$ half-space. $x^{2}=0$ gives the
 boundary and we consider the supersymmetric boundary conditions on this.}
\label{figurea0}
\end{center}
\end{figure}

In this paper, we focus on 1/2 BPS boundary conditions of $\Ncal=2$
supersymmetric theories in three dimensions.  There are two
possibilities of the preserved supersymmetry (SUSY), $\Ncal=(1,1)$ and $\Ncal=(2,0)$, as classified in \cite{Berman:2009kj}.  We call them ``A-type'' and ``B-type'' respectively in this paper since they are analogs of the A-type and B-type boundary conditions in two-dimensional $\Ncal=(2,2)$ theories.
We consider supersymmetric field theories on the flat half spacetime $\mathbb{R}^{1,1}\times
\mathbb{R}_{+}$ for simplicity. 
Here $\mathbb{R}^{1,1}$ is parametrized by the time coordinate $x^{0}$
and the spatial coordinate $x^{1}$. $\mathbb{R}_{+}$ is an infinite
half line $x^{2}\ge 0$ (see Figure \ref{figurea0}).   Our way to examine boundary conditions 
is to check whether the component of the SUSY current orthogonal to the boundary $J^2$ vanishes, as done in \cite{Gaiotto:2008sa}.

Let us summarize the results in this paper.  We first consider the
$\mathcal{N}=2$ Landau-Ginzburg theory. We employ the brane picture in
the same way as in the two-dimensional case. %In this case, a brane%
%modify
In some sense, our brane is understood as the extended objects upon
which membranes can end
%modify
\footnote{Here we use the term ``membrane'' in a broad meaning. 
We do not claim that our membrane is the same as that considered in M-theory.}.
 An A-type brane (A-brane) is a Lagrangian submanifold on which the imaginary part of the superpotential is constant,  while a B-type brane (B-brane) is a holomorphic submanifold 
on which the superpotential is constant. 

Second we explore 1/2 BPS boundary conditions in $\Ncal=2$ pure Maxwell theory.  This theory is dual to a free field theory which contains a chiral multiplet.  We find a few classes of A-type and B-type boundary conditions and interpret them in the free chiral multiplet theory. The results are completely consistent with the analysis of the Landau-Ginzburg theory.

Finally we consider 1/2 BPS boundary conditions in $\Ncal=2$
supersymmetric quantum electrodynamics (SQED).  This theory is supposed
to be dual to a Landau-Ginzburg model called ``the XYZ model''
\cite{Aharony:1997bx, deBoer:1997kr}.  We conjecture an example of
mirror symmetry with boundary and give an evidence in the picture of the
moduli space.

%modify
Although we find the nontrivial solutions of supersymmetric boundary conditions in
three-dimensional $\mathcal{N}=2$ theories, these are not complete. 
In the two-dimensional contexts, it was discussed that we may also couple
bulk theories to boundary degrees of freedom by introducing massless
vector bosons, Chan-Paton spaces and so on \cite{Herbst:2008jq}. 
To complete our analysis, we should take three-dimensional analogs into
account. They are extremely interesting, but will be deferred to future
work.

The organization of this paper is as follows.
In section 2 we determine the supersymmetric boundary conditions for
three-dimensional $\mathcal{N}=2$ Landau-Ginzburg model. 
In section 3 we derive the supersymmetric boundary conditions for three-dimensional $\mathcal{N}=2$ pure Maxwell theory. 
Then  we consider the duality between pure Maxwell theory and chiral matter theory. 
In section 4 we also present the supersymmetric boundary conditions
in 3-dimensional $\mathcal{N}=2$ SQED. Finally section 5 concludes with a
discussion of the relating problems and future works. 
The Appendixes contain our notations and some useful formulae in
three-dimensional $\mathcal{N}=2$ field theories.

\section{Landau-Ginzburg model}\label{sec:LG}
In this section we discuss Landau-Ginzburg models in three dimensions. 
We find supersymmetric boundary conditions that preserve half of the
supersymmetry and show that the subspace of the sigma model arises as
Lagrangian submanifolds or holomorphic submanifolds.

%sec1

Let us consider the Landau-Ginzburg model which has $n$ chiral
superfields $\Phi^{i} (i=1,\cdots,n)$.  See the Appendixes for the detail of the convention.
The Lagrangian is
\begin{equation}
 \mathcal{L}
=K(\Phi, \bar{\Phi})|_{-\theta\theta\thetab\thetab}
+W(\Phi)|_{\theta\theta}
+\bar{W}(\bar{\Phi})|_{-\thetab\thetab}. 
\end{equation}
Here $K(\Phi, \bar{\Phi})$
 is the K\"{a}hler potential 
 and $W(\Phi)$ is the superpotential.

The K\"{a}hler potential term is expressed in component fields as
\begin{align}
K|_{-\theta\theta\thetab\thetab}
=&K_{i\bar{j}}F^{i}\bar{F}^{\bar{j}}+\frac12 K_{i\bar{j}\bar{k}}
F^{i}(\psib^{\bar{j}}\psib^{\bar{k}})
-\frac12K_{ij\bar{k}}
\bar{F}^{\bar{k}}(\psi^{i}\psi^{j}) 
 -\frac{1}{4}K_{ij\bar{k}\bar{l}}(\psi^{i}\psi^{j}\psib^{\bar k}\psib^{\bar l})
\nonumber \\&
 -K_{i\bar{j}}\partial_{\mu}\phi^{i}\partial^{\mu}\phib^{\bar{j}}
-\frac{i}{2}K_{i\bar{j}}\psib^{\bar{j}}\sigma^{\mu}\partial_{\mu}\psi^{i}
-\frac{i}{2}K_{i\bar{j}}\psi^{i}\sigma^{\mu}\partial_{\mu}\psib^{\bar{j}}
 \nonumber \\
&-\frac{i}{2}K_{ij\bar{k}}(\partial_{\mu}\phi^{i})(\psib^{\bar{k}}\sigma^{\mu}\psi^{j})+\frac{i}{2}K_{ij\bar{k}}(\partial_{\mu}\phib^{\bar{j}})
(\psib^{\bar{k}}\sigma^{\mu}\psi^{i}),
\end{align}
where we use the abbreviation $K_{i\bar{j}}:=\frac{\partial^{2}
K}{\partial \phi^{i}\partial \phib^{\bar{j}}}$ and so on.

On the other hand, the contribution from superpotential is
\begin{align}
W|_{\theta\theta}+\bar{W}|_{-\thetab\thetab}
=F^{i}W_{i}-\frac{1}{2}(\psi^{i}\psi^{j})W_{ij}+(c.c.) ,
\end{align}
where $(c.c.)$ denotes the complex conjugation. We also use the abbreviations $W_{i}:=\frac{\partial W}{\partial \phi^{i}}, W_{ij}:=\frac{\partial^2 W}{\partial \phi^{i}\partial \phi^{j}}$.

The supersymmetry transformation of this system is expressed as
\begin{align}
\delta\phi^{i}&=\sqrt{2}\epsilon\psi^{i},  \\
\delta\psi^{i}&=\sqrt{2}i\gamma^{\mu}\bar{\epsilon}
\partial_{\mu}\phi^{i}+\sqrt{2}\epsilon F^{i} ,\\
\delta F^{i}&=\sqrt{2}i\bar{\epsilon}\sigma^{\mu}\partial_{\mu}\psi^{i},
\end{align}
\begin{align}
\delta\phib^{\bar i}&=-\sqrt{2}\bar{\epsilon}\psib^{\bar i} ,\\
\delta\psib^{\bar i}&=-\sqrt{2}i\gamma^{\mu}\epsilon\partial_{\mu}\phib^{\bar i}+\sqrt{2}\bar{\epsilon}\bar{F}^{\bar i}
 ,\\
\delta\bar{F}^{\bar i}&=\sqrt{2}i\epsilon\sigma^{\mu}\partial_{\mu}\psib^{\bar i}.
\end{align}
We can calculate the supercurrents
\footnote{An improvement transformation may give rise to some ambiguity to
determine the supercurrents\cite{Dumitrescu:2011iu}. Understanding their
effects may create an interesting problem. We thank Yu Nakayama for discussions on these points.}
\begin{align}
J^{\mu}
&=-\sqrt{2}K_{i\bar{j}}(\partial^{\mu}\phib^{\bar{j}})\psi^{i}
+\sqrt{2}K_{i\bar{j}}(\partial_{\nu}\phib^{\bar{j}})\gamma^{\mu\nu}\psi^{i}-\sqrt{2}i\gamma^{\mu}\psib^{\bar{i}}\bar{W}_{\bar{i}},
 \nonumber \\
\bar{J}^{\mu}
&=-\sqrt{2}K_{i\bar{j}}(\partial^{\mu}\phi^{i})\psib^{\bar{j}}
+\sqrt{2}K_{i\bar{j}}(\partial_{\nu}\phi^{i})
\gamma^{\mu\nu}\psib^{\bar{j}}
+\sqrt{2}i\gamma^{\mu}\psi^{i}W_{i}.\label{LGcurrent}
\end{align}

Here we investigate this system in the half-space $x^{2}\ge 0$.  We
restrict ourselves to the case without boundary terms or boundary
degrees of freedom almost throughout this paper.  Then the equations of
motion give nontrivial constraints on the boundary conditions. Let us
use the capital label $I$ which takes values of both $i$ and $\bar{i}$.  The target space metric $g_{IJ}$ is defined as
\begin{align}
g_{i\bar{j}}=g_{\bar{j}i}=K_{i\bar{j}},\qquad g_{ij}=g_{\bar{i}\bar{j}}=0.
\end{align}
The boundary term coming from the bosonic term becomes
\begin{align}
\delta S_{B, bdy}=\int d^2x  g_{IJ}\delta \phi^{I} \del_{2}\phi^{J}. \label{BT0}
\end{align}
We can use the target space brane picture in the same way as the string theory. See Figures.~\ref{figurea5} and \ref{figurea6}.  The target space vector $\delta \phi^I$ is tangent to the brane by definition.  We should impose the boundary condition in which the boundary term \eqref{BT0} vanishes for an arbitrary tangent 
vector $\delta \phi^{I}$.  Thus the target space vector $\del_{2}\phi^{I}$ is normal to the brane.

On the other hand the fermionic boundary term becomes
\begin{align}
\delta S_{F,bdy}=\int d^2x \frac{i}{2}g_{IJ}\delta\psi^{I}\sigma^{2}\psi^{J}.\label{BT1}
\end{align}
We impose the boundary condition
\begin{align}
\gamma^2\psi^{I}=S^{I}{}_{J}\psi^{J} \label{defS0}
\end{align}
with a $\phi$ dependent matrix $S^{I}{}_{J}$.  $(\gamma^2)^2=1$ leads to the constraint
\begin{align}
S^{I}{}_{J}S^{J}{}_{K}=\delta^{I}_{K}.\label{SS}
\end{align}
We require the boundary term \eqref{BT1} to vanish.  Then another constraint on $S^{I}{}_{J}$ is obtained
\begin{align}
g_{IJ}S^{I}{}_{K}S^{J}{}_{L}=g_{KL}. \label{SgS=g}
\end{align}

Let us turn to the supersymmetry of the boundary condition. 
A boundary condition preserves supersymmetry if and only if the
component of the SUSY current normal to the boundary vanishes. 
Thus supersymmetric boundary condition satisfies
\begin{align}\label{susycond1}
 0=\epsilon J^{2}-\bar{\epsilon}\bar{J}^{2},
\end{align}
for a certain class of $\epsilon$.
There are two kinds of choices of $\epsilon$ for 1/2 BPS boundary as considered in
\cite{Berman:2009kj}
\begin{align}\label{absusy1}
&(A)\quad \gamma^{2}\epsilon=\bar{\epsilon},\quad \Ncal=(1,1) \text{ type}, \nonumber \\
&(B)\quad \gamma^{2}\epsilon=\epsilon,\quad \Ncal=(2,0) \text{ type}. 
\end{align}
We call them A-type and B-type, respectively, in this paper.  They are actually analogous to the A-type and B-type boundary conditions in two-dimensional $\Ncal=(2,2)$ theories.

With the SUSY currents expressions \eqref{LGcurrent}, the SUSY condition \eqref{susycond1} becomes
\begin{align}
\label{susycondkw}
0&=-K_{i\bar{j}}(\partial^{2}\phib^{\bar{j}})(\epsilon\psi^{i})
+K_{i\bar{j}}(\partial_{\nu}\phib^{\bar{j}})(\epsilon
 C\gamma^{2\nu}\psi^{i})
-i(\epsilon\sigma^{2}\bar{\psi}^{\bar{i}})\bar{W}_{\bar{i}} \nonumber \\
&\quad
 +K_{i\bar{j}}(\partial^{2}\phi^{i})(\bar{\epsilon}\psib^{\bar{j}})
-K_{i\bar{j}}(\partial_{\nu}\phi^{i})(\bar{\epsilon}C\gamma^{2\nu}\psib^{\bar{j}})
-i(\bar{\epsilon}\sigma^{2}\psi^{i})W_{i}.
\end{align}
Let us see the geometric meaning of this condition for A-type and B-type.

\subsection{A-type $\gamma^{2}\epsilon=\bar{\epsilon}$}
Now we want to discuss the brane of an A-type boundary condition. 
We call it ``A-brane.'' 
Here we will show that an A-brane is a Lagrangian submanifold on which $\im{W}$ is constant.  This result is similar to the two-dimensional case \cite{Hori:2000ck}.

It is natural to employ the ansatz for the boundary condition for the fermions
\begin{align}
\gamma^{2}\psi^{i}={S^{i}}_{\bar{j}}\psib^{\bar{j}}.
\end{align}
Here $S^{i}{}_{\bar{j}}$ is a $\phi$ dependent matrix. 
Then the matrix $S^{I}{}_{J}$ in Eq.~\eqref{defS0} becomes
\begin{align}
S^{I}{}_{J}=\left(
\begin{array}{cc}
0&{S^{*\bar{i}}}_{j}\\
{S^{i}}_{\bar{j}}&0\\
\end{array}
\right).\label{defS}
\end{align}

\begin{figure}
\begin{center}
\includegraphics[width=8cm]{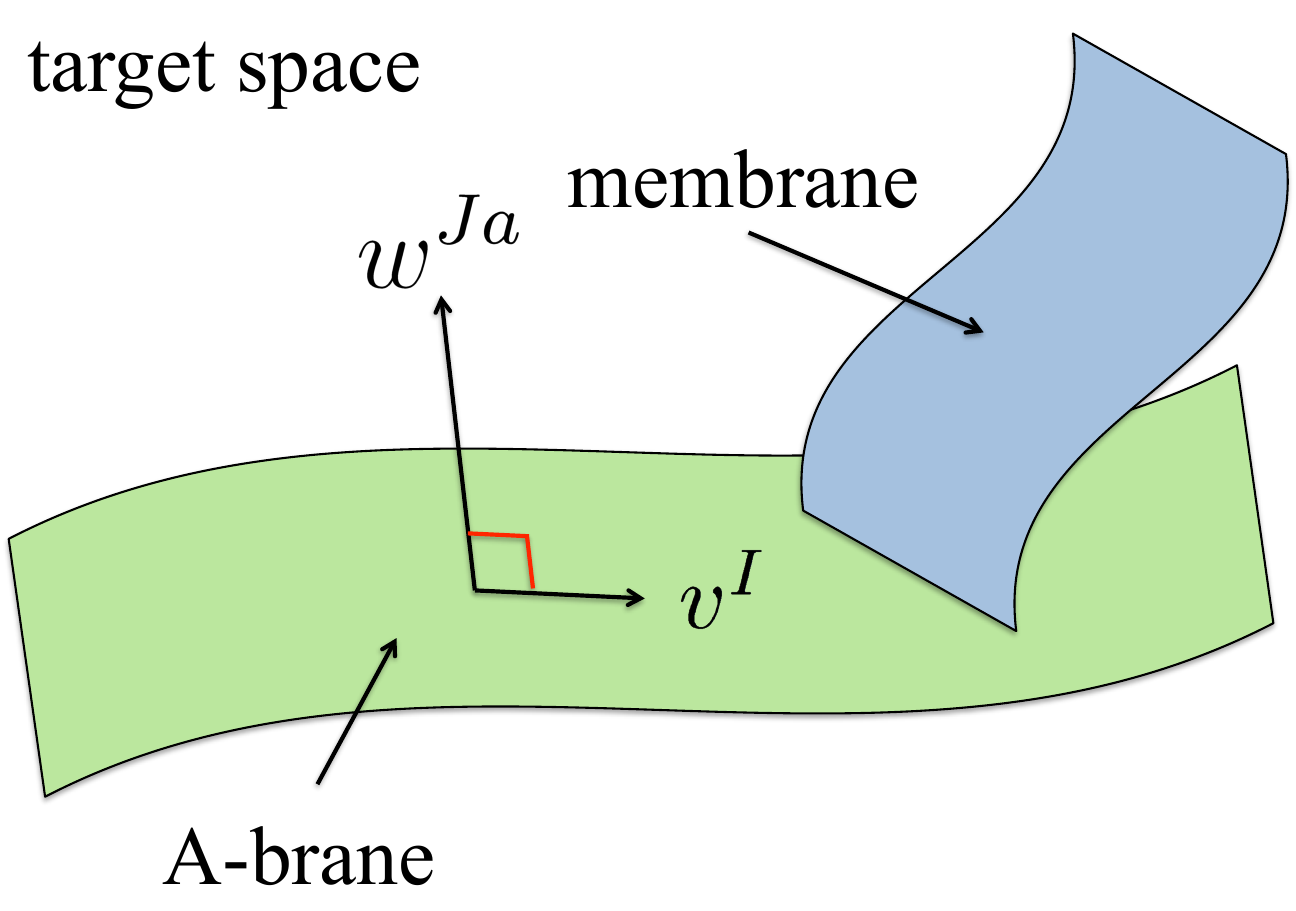}
\caption{Membrane, shown in blue, ending on the A-brane, shown in
 green. They live  in the target space. $v^{I}$ is
parallel to the tangent direction and $w^{Ja}$ is in the normal
 direction of the A-brane.}
\label{figurea5}
\end{center}
\end{figure}

Before going,  we introduce 
some useful expressions.
We introduce the target space vectors $v^{I}$ and $w^{Ia},\ a=0,1$ as
\begin{align}
v^{I}:=\begin{cases}
\epsilon\psi^{i}&(I=i) \cr
-\bar{\epsilon}\psib^{\bar{i}}&(I=\bar{i}) \cr
\end{cases}
\qquad
w^{Ia}:=\begin{cases}
\bar{\epsilon}\sigma^{a}\psi^{i}&(I=i) \cr
-\epsilon\sigma^{a}\psib^{\bar{i}}&(I=\bar{i})\cr
\end{cases}. \label{defvw}
\end{align}
Then the condition (\ref{susycondkw}) is rewritten as
\begin{align}
0=-g_{IJ}\partial_{2}\phi^{I}v^{J}
-g_{IJ}\partial_{a}\phi^{I}w^{Ja}
+iW_{i}v^{i}-i\bar{W}_{\bar{i}}\bar{v}^{\bar{i}}, \qquad \textrm{where}\quad a=0,1.
\end{align}
The above condition is satisfied by the ansatz\footnote{We could not
find any other solutions without boundary terms, although we could not
prove this is the only solution.  If some boundary terms are included,
we may have other solutions.}: 
\begin{align}
g_{IJ}\partial_{2}\phi^{I}v^{J}&=0 ,\\
g_{IJ}\partial_{a}\phi^{I}w^{Ja}&=0 ,\\
iW_{i}v^{i}-i\bar{W}_{\bar{i}}\bar{v}^{\bar{i}}&=0, \qquad \textrm{where}\quad a=0,1.
\end{align}
Since $\del_2\phi^{I}$ is normal to the brane and $\del_a \phi^{I}$ is tangent to the brane, first two conditions imply that $v^{I}$ is tangent to the brane and that $w^{Ia}$ is normal to the brane respectively as shown in Fig.~\ref{figurea5}.
The third condition implies that the imaginary part of the superpotential
      $\textrm{Im}W$ is constant on the brane.

Let us define the K\"ahler form of the target space $\omega_{IJ}$
\begin{align}
\omega_{IJ}:= \begin{cases} 
\omega_{i\bar{j}}=iK_{i\bar{j}} & (I=i, J=\bar{j})\cr
\omega_{\bar{i}j}=-iK_{\bar{i}j} & (I=\bar{i}, J=j)\cr
0 & (\text{otherwise}) 
\end{cases} 
\end{align}
In order to show that the brane is a Lagrangian submanifold, we should check that
\begin{enumerate}
\item The real dimension of the submanifold is $n$ in the complex $n$-dimensional target space.
\item For two arbitrary tangent vectors $v^{I}$ and $v'^{I}$
\begin{align}
\label{lagrangiancond}
\omega_{IJ}v^{I}v^{'J}=0,
\end{align}
are satisfied.
\end{enumerate}
Let us check these propositions one by one.

First, notice that from the definition \eqref{defvw},  
\begin{align}
&{S^{I}}_{J}v^{J}=v^{J}, \nonumber \\
&{S^{I}}_{J}w^{Ja}=-w^{Ja}
\end{align}
are satisfied. In other words tangent vectors and normal vectors are eigen vectors of $S$ with eigenvalues $1$ and $-1$ respectively.  On the other hand  $S^2=1$ and $\Tr(S)=0$  are satisfied because of Eqs.~\eqref{SS}, \eqref{defS} respectively.  Thus the $(2n)\times (2n)$ matrix $S$ has real $n$-dimensional eigenspace with eigenvalue $+1$.  In other words the A-brane is real $n$ dimensions.

The second one is shown as follows.  Eq.~\eqref{SgS=g} is written in this case as
\begin{align}
 K_{i\bar{j}}{S^{i}}_{\bar{k}}{S^{*\bar{j}}}_{l}
=K_{l\bar{k}}.
\end{align}
By using this relation the left-hand side of the relation \eqref{lagrangiancond} is rewritten as
\begin{align}
\label{lagrangiancond2}
\omega_{IJ}v^{I}v'^{J}= i\left(K_{i\bar{j}}v^{i}\bar{v}^{'\bar{j}}-K_{\bar{i}j}({{S^{*}}^{\bar{i}}}_{k}v^{k})({S^{j}}_{\bar{l}}\bar{v}^{'\bar{l}})\right)=0.
\end{align}
This is the relation \eqref{lagrangiancond}.  As a result we have shown that an A-brane is a Lagrangian submanifold.
%Again fermionic boundary terms give us useful information. 
%Let us take the variation of $S_{F}$ in terms of
%$\psi$. Using the relations in A-type boundary conditions we
%obtain the following boundary term:
%\begin{align}
%\delta S_{F}|_{\textrm{boundary}}
%&=iK_{i\bar{j}}\psib^{\bar{j}}\sigma^{2}\delta\psi^{i}+
%iK_{i\bar{j}}\psi^{i}\sigma^{2}\delta\psib^{\bar{j}} \nonumber \\
%&=i\psib S^{T}(S^{*T}KS-K)\delta\psib.
%\end{align}
%This is required to vanish by action principle, which requires that
%\begin{align}
% K_{i\bar{j}}{S^{i}}_{\bar{k}}{S^{*\bar{j}}}_{l}
%=K_{l\bar{k}}.
%\end{align}
%Therefore (\ref{lagrangiancond2}) is satisfied and target space is
%Lagrangian if we choose A-type
%supersymmetric boundary conditions.

\subsection{B-type $\gamma^{2}\epsilon=\epsilon$}
\begin{figure}
\begin{center}
\includegraphics[width=8cm]{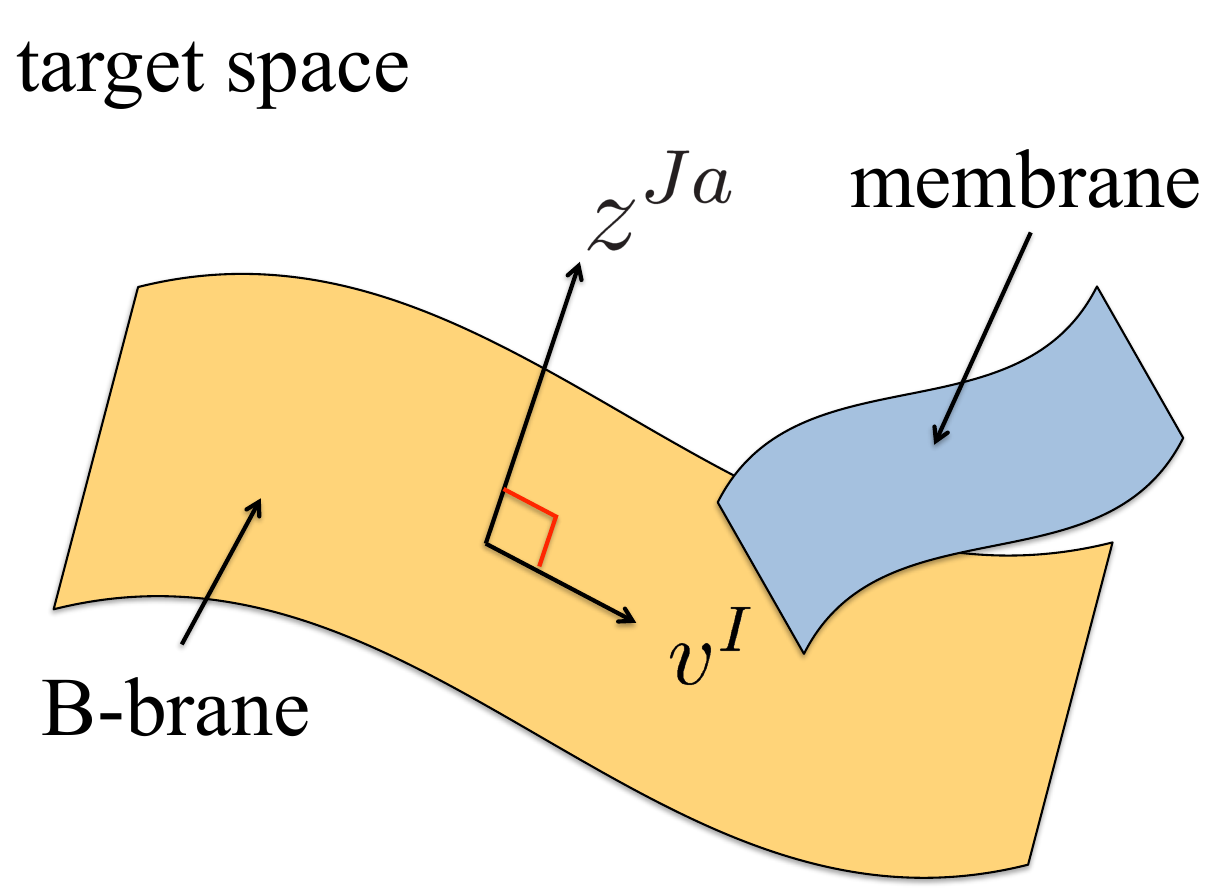}
\caption{Membrane, shown in blue  ending on the B-brane, shown in orange. They live in
 the target space. $v^{I}$ is in the tangent direction and $z^{Ja}$ is in the
 normal direction of B-brane.}
\label{figurea6}
\end{center}
\end{figure}

Next let us turn to the brane of B-type boundary condition.  We call it ``B-brane'' 
(see Figure \ref{figurea6}).
We will show that a B-brane is a holomorphic submanifold on which superpotential $W$ is constant.  This is also similar to the two-dimensional B-type boundary condition\cite{Hori:2000ck}.

It is natural to put the ansatz for the boundary condition for the fermions
\begin{align}
\gamma^{2}\psi^{i}={R^{i}}_{j}\psi^{j},
\end{align}
where $R^{i}{}_{j}$ is a $\phi$ dependent matrix.  Then $S^{I}{}_{J}$ in Eq.~\eqref{defS0} becomes
\begin{align}
S^{I}{}_{J}=\left(
\begin{array}{cc}
R^{i}{}_{j} & 0\\
0 & {{R^*}^{\bar{i}}}_{\bar{j}}\\
\end{array}
\right).\label{defR}
\end{align}

Here we define $u^{I},\ z^{Ia},\ (a=0,1)$ as
\begin{align}
u^{I}:=\begin{cases}
i\bar{\epsilon}\psi^{i}&(I=i)\cr
i\epsilon\psib^{\bar{i}}&(I=\bar{i})\cr
\end{cases}
\qquad
z^{Ia}:=
\begin{cases}
\epsilon\sigma^{a}\psi^{i}&(I=i)\cr
-\bar{\epsilon}\sigma^{a}\psib^{\bar{i}}&(I=\bar{i})\cr
\end{cases}. \label{defuz}
\end{align}

Then the supersymmetric boundary condition is rewritten as
\begin{align}
0=-g_{IJ}(\partial^{2}\phi^{I})v^{J}
-g_{IJ}(\partial_{a}\phi^{I})z^{Ja}
-i(\epsilon\psib^{\bar{i}}){(R^{*})^{\bar{i}}}_{\bar{j}}\bar{W}_{\bar{i}}
-i(\bar{\epsilon}\psi^{j}){R^{i}}_{j}W_{i},
\end{align}
which is satisfied by the ansatz
\begin{align}
g_{IJ}\partial_{2}\phi^{I}v^{J}&=0 ,\\
g_{IJ}\partial_{a}\phi^{I}z^{Ja}&=0 ,\qquad
 \mathrm{where} \quad a=0,1,\\
u^{i}W_{i}+\bar{u}^{\bar{i}}\bar{W}_{\bar{i}}&=0 .\label{BWcond}
\end{align}
The first condition and second one imply that the target space vector $v^{I}$ is tangent to the brane and that $z^{Ia}$ is normal to the brane respectively.

Now we would like to show that a B-brane is a holomorphic submanifold.
It is necessary and sufficient to show
\begin{align}
\label{holcond}
\omega_{IJ}v^{I}z^{Ja}=0
\end{align}
for an arbitrary tangent vector $v^{I}$ and an arbitrary normal vector $z^{Ja}$. From the definitions \eqref{defvw} and \eqref{defuz}, the relations
\begin{align}
v^{I}&=-{S^{I}}_{J}v^{J}, \\
z^{Ia}&={S^{I}}_{J}z^{Ja} \label{Beigenvectors}
\end{align}
are obtained.
In other words, the tangent space is the eigenspace of $S$ with eigenvalue $-1$, and the normal space is that with eigenvalue $+1$.  The relation \eqref{SgS=g} reads in this B-type ansatz
\begin{align}
 K_{i\bar{j}}{R^{*\bar{j}}}_{\bar{k}}{R^{i}}_{l}=K_{\bar{k}l}.
\end{align}
By using this equation and Eq.~\eqref{Beigenvectors} the left-hand side of Eq.~\eqref{holcond} is rewritten as
\begin{align}
\label{holcond2}
i(-K_{i\bar{j}}{R^{i}}_{k}v^{k}{R^{*\bar{j}}}_{\bar{l}}\bar{z}^{\bar{l}a}
-K_{\bar{i}j}{R^{*\bar{i}}}_{\bar{k}}
\bar{v}^{\bar{k}}{R^{j}}_{l}z^{la})=0.
\end{align}
Thus the relation \eqref{holcond} is satisfied and we can conclude that the B-brane is a holomorphic submanifold.

Next let us turn to show that the superpotential is constant on the B-brane.  The target space vector $u^{I}$, defined in Eq.~\eqref{defuz} satisfies
\begin{align}
S^{I}{}_{J}u^{J}=-u^{I}.
\end{align}
In other words $u^{I}$ is a eigenvector with eigenvalue $-1$ and therefore $u^{I}$ is a tangent vector.  As a result Eq.~\eqref{BWcond} implies $W$ is constant on the B-brane.

%sec3

\section{Pure Maxwell theory}
\subsection{Review of the abelian duality}
In this section we study three-dimensional $\mathcal{N}=2$ pure Abelian gauge theory.
The supersymmetric Lagrangian of the vector multiplet is given by
\begin{align}
 \mathcal{L}_{\mathrm{gauge}}
&=-\frac{1}{e^{2}}\Sigma^{2}|_{-\theta\theta\thetab\thetab} \nonumber \\
&=\frac{1}{e^{2}}
\left(
-\frac{1}{4}F_{\mu\nu}F^{\mu\nu}
-i\bar{\lambda}\sigma^{\mu}\partial_{\mu}\lambda
-\frac{1}{2}\partial^{\mu}\sigma\partial_{\mu}\sigma
+\frac{1}{2}D^{2}
\right),
\end{align}
where we denote the gauge coupling constant by $e$. 
$\Sigma$ is the linear multiplet defined by
$\Sigma=-\frac{i}{2}\bar{D}DV$. For the detail of the convention, see the Appendixes.
%The lowest component of $\Sigma$ is the
%scalar $\sigma$. 

Let us first review the duality between pure Abelian gauge theory and massless free theory.
As discussed in \cite{deBoer:1997kr}, let us start with the action
\begin{align}
\label{action}
S=\int d^{3}x \int d^{4}\theta
\left(
-\frac{1}{e^{2}}\Sigma^{2}+\Sigma(\Phi+\bar{\Phi})
\right),
\end{align}
where $\Sigma$ is a general real superfield and $\Phi$ is a chiral superfield. The fermion integral $\int d^4\theta$ picks up the coefficient of $(-\theta\theta\bar{\theta}\bar{\theta})$.
 Two dual theories may be thought of as two choices of variables in path
integral. 
In other words, we can interpret such dualities as Legendre transformations.

If we integrate out $\Phi$ and $\bar{\Phi}$, 
we obtain constraints $\bar{D}^{2}\Sigma=0$ and 
$D^{2}\Sigma=0$, which mean that $\Sigma$ is a linear multiplet.
 Thus this leads to pure Maxwell theory if we integrate out $\Phi$
       and $\bar{\Phi}$.

On the other hand, we can integrate out $\Sigma$ first. 
This integral can be performed by solving the equation of motion for $\Sigma$ and substituting $\Sigma$ in the action \eqref{action} with this classical solution.
The classical equation of motion is given by
\begin{align}
\label{dualtrans}
\Sigma
=\frac{e^{2}}{2}(\Phi+\bar{\Phi}).
\end{align}
By substituting this in the action, we can rewrite the action as a
      functional of $\Phi$ and $\bar{\Phi}$:
\begin{align}
S=\frac{e^{2}}{2}\int d^{3}x \int d^{4}\theta \bar{\Phi}\Phi
\end{align}
Now this gives chiral matter theory characterized by the K\"{a}hler
      potential $K=\frac{e^{2}}{2}\bar{\Phi}\Phi$.

Let us see the duality transformation in components. 
 By expanding (\ref{dualtrans}), we obtain the following dictionary:
\begin{align}
\sigma&=e^{2}(\textrm{Re}\phi) \\
G_{\mu}&:=\frac12 \epsilon_{\mu\nu\rho}F^{\nu\rho}=e^{2}\partial_{\mu}(\textrm{Im}\phi) =: \partial_{\mu}\rho\label{dual-photon}\\
\lambda&=\frac{e^{2}}{\sqrt{2}}\psib \\
\bar{\lambda}&=\frac{e^{2}}{\sqrt{2}}\psi.
\end{align}
The relation \eqref{dual-photon} shows that $\im \phi$ is the dual photon. 
It is also convenient to define $\rho:=e^2\im\phi$. Then $\sigma+i\rho$ is the holomorphic coordinate.  From charge quantization, we see that $\rho$ is
periodic:
\begin{align}
\rho&\sim \rho+\frac{e^{2}}{2}.
\end{align}

\subsection{Supersymmetric boundary conditions}

In Wess-Zumino gauge, the SUSY transformation for the component fields of the vector multiplet are
\begin{align}
\delta
 A_{\mu}&=i\overline{\epsilon}\sigma_{\mu}\lambda+i\epsilon\sigma_{\mu}\bar{\lambda},
 \nonumber \\
\delta \sigma&=\epsilon\bar{\lambda}-\bar{\epsilon}\lambda, \nonumber \\
\delta
 \bar{\lambda}&=-i\bar{\epsilon}D-\frac{1}{2}\gamma^{\mu\nu}\bar{\epsilon}F_{\mu\nu}+i\gamma^{\mu}\bar{\epsilon}\partial_{\mu}\sigma,
 \nonumber \\
\delta \lambda
&=i\epsilon D-\frac{1}{2}\gamma^{\mu\nu}\epsilon F_{\mu\nu}
-i\gamma^{\mu}\epsilon \partial_{\mu}\sigma, \nonumber \\
\delta D
&=-\epsilon\sigma^{\mu}\partial_{\mu}\bar{\lambda}
+\bar{\epsilon}\sigma^{\mu}\partial_{\mu}\lambda.
\end{align}

Given the above Lagrangian and the SUSY transformation, we can 
calculate supercurrents as
\begin{align}
&J^{\mu}
=-iF^{\mu\nu} \gamma_{\nu}\bar{\lambda}
+\frac{i}{2}\epsilon^{\mu\rho\sigma}\bar{\lambda} F_{\rho\sigma}
+\gamma^{\mu\nu}\bar{\lambda} \partial_{\nu}\sigma
-\bar{\lambda}
\partial^{\mu}\sigma, \nonumber \\
&\bar{J}^{\mu}
=+iF^{\mu\nu}\gamma_{\nu}\lambda
-\frac{i}{2}
\epsilon^{\mu\rho\sigma}\lambda F_{\rho\sigma}
+\gamma^{\mu\nu}\lambda \partial_{\nu}\sigma
-\lambda \partial^{\mu}\sigma.
\end{align}
Then the supersymmetric boundary condition for vector multiplet is given by
\begin{align}
\label{purecond1}
0&=\epsilon J^{2}-\bar{\epsilon}\bar{J}^{2} \nonumber \\
&=-iF^{2a}(\epsilon\sigma_{a}\bar{\lambda})+i(\epsilon\bar{\lambda})F_{01}
+(\epsilon C\gamma^{2a}\bar{\lambda})\partial_{a}\sigma
-(\epsilon\bar{\lambda})\partial_{2}\sigma \nonumber \\
&\quad -iF^{2a}(\bar{\epsilon}\sigma_{a}\lambda)
+i(\bar{\epsilon}\lambda)F_{01}
-(\bar{\epsilon}C\gamma^{2a}\lambda)\partial_{a}\sigma
+(\bar{\epsilon}\lambda)\partial_{2}\sigma, \qquad \textrm{where} \quad a=0,1.
\end{align}

%We assume two types of the solutions:
%\begin{align}
%&\gamma^{2}\lambda=e^{i\alpha}\bar{\lambda} \nonumber \\
%&\gamma^{2}\lambda=e^{i\beta}\lambda,
%\end{align}
%which lead us to consider the following combinations:
%\begin{align}\label{absusy}
%&(A)\quad \gamma^{2}\epsilon=\bar{\epsilon}, \quad\gamma^{2}\lambda=e^{i\alpha}\bar{\lambda} \nonumber \\
%&(B)\quad \gamma^{2}\epsilon=\epsilon, \quad\gamma^{2}\lambda=e^{i\beta}\lambda.
%\end{align}

We focus on the case without boundary terms or boundary degrees of freedom in this paper, except the boundary theta term:
\begin{align}
\label{thetaterm1}
S_{\vartheta}=\frac{\vartheta}{2\pi}\int_{x^2=0}
dx^{0}dx^{1}F_{01}.
\end{align} 
This theta term corresponds to the shift for the value of dual photon at the boundary in the dual picture as pointed out in \cite{Kapustin:2009av}.  In order to see this we begin by defining the reference point $\rho_{0}$ as
\begin{align}
\rho_{0}:=\langle \rho(x)\rangle_{0}=\int D\rho\rho(x)e^{iS_{0}},
\end{align}
where 
\begin{align}
S_{0}=-\frac{1}{2e^{2}}\int
d^{3}x \partial_{\mu}\rho\partial^{\mu}\rho,
\end{align}
which is the kinetic term of $\rho$ determined by (\ref{dualtrans}).
This boundary term (\ref{thetaterm1}) is written in terms of the dual photon $\rho$:
\begin{align}
S_{\vartheta}=\frac{\vartheta}{2\pi}
\int dx^{0}dx^{1}\partial_{2}\rho.
\end{align}
Using $S_{0}+S_{\vartheta}$ instead of $S_{0}$, we obtain
\begin{align}
\rho_{\vartheta}
:=\langle \rho(x)\rangle_{\vartheta}
%=\int D\rho\rho(x)e^{i(S_{0}+S_{\theta})}
=\rho_{0}+\frac{e^{2}}{4\pi}\vartheta.
\end{align}
{}From this result we can also see the periodicity of $\rho$ by $\vartheta$ whose domain is
$0\le \vartheta <2\pi$.  This correspondence is also explained by the usual dualization procedure including appropriate boundary terms.  In two dimensions with boundary this have been done in \cite{Hori:2000ic} and it is straightforward to extend it to three dimensions.\footnote{We would like to thank Kentaro Hori for explanation.}

This Abelian duality is an analog of T-duality in two-dimensions.  For example the boundary theta term is the analog of the Wilson line in two-dimensions whose dual is the position of the D-brane.  One may think that this three-dimensional abelian duality exchanges an A-brane and a B-brane from the analogy of two-dimensions. However it is not true.  An A-brane and a B-brane in three-dimensions 
preserve different type of supersymmetry. They are $\Ncal=(1,1)$ and $\Ncal=(2,0)$ on the boundary respectively.
Thus an A-brane and a B-brane cannot be dual to each other.  The dual of an A-brane is an A-brane, and that of a B-brane is a B-brane. 

We consider in this paper several examples of boundary conditions with simple ansatze and see the duality, instead of the general classification of the boundary conditions.  Let us examine the A-type ($\Ncal=(1,1)$ type) and the B-type ($\Ncal=(2,0)$ type)  \cite{Berman:2009kj} boundary conditions.

\subsubsection{A-type $\gamma^{2}\epsilon=\bar{\epsilon}$}
We put the ansatz
\begin{align}
\gamma^{2}\lambda=e^{i\alpha}\bar{\lambda}
\end{align}
with a real parameter $\alpha$. Then the condition (\ref{purecond1}) is rewritten as
\begin{align}
0
&=2ie^{-\frac{i\alpha}{2}}\epsilon\sigma^{a}\lambda
\left(
\cos\frac{\alpha}{2} G_{a}+\sin\frac{\alpha}{2}\partial_{a}\sigma
\right) \nonumber \\
&\quad+2e^{-\frac{i\alpha}{2}}\epsilon
\sigma^{2}\lambda\left(
\sin\frac{\alpha}{2} G_{2}-\cos\frac{\alpha}{2} \partial_{2}\sigma
\right),\qquad \textrm{where} \quad a=0,1,
\end{align}
where $G^{\mu}$ is the dual strength of the gauge field defined by $G^{\mu}:=\frac{1}{2}\epsilon^{\mu\rho\eta}F_{\rho\eta}$.
To satisfy the above condition, we should require that
\begin{align}
\label{maxwella}
\cos\frac{\alpha}{2}G_{a}
+\sin\frac{\alpha}{2}\partial_{a}\sigma&=0, \nonumber \\
\sin\frac{\alpha}{2}G_{2}-\cos\frac{\alpha}{2}\partial_{2}\sigma&=0,
 \qquad  \textrm{where} \quad a=0,1.
\end{align}
Note that we have theta term contribution \eqref{thetaterm1} except when $\alpha=\pi$.

\begin{figure}[H]
\begin{center}
\subfigure[$\alpha=0$]{\includegraphics*[width=.3\linewidth]{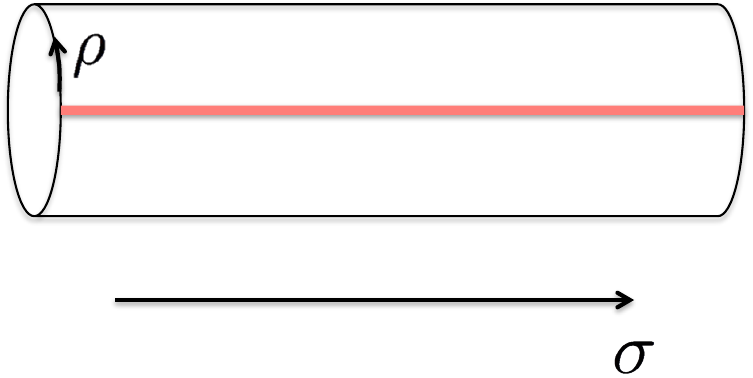}
\label{figurea7a}}
\subfigure[$\alpha=\pi$]{\includegraphics*[width=.3\linewidth]{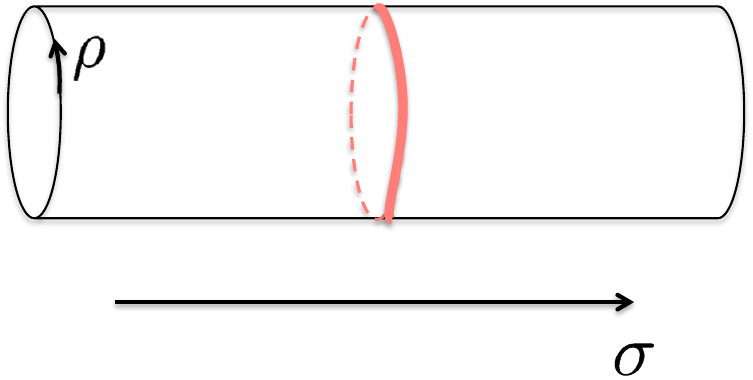}
\label{figurea7b}}
\subfigure[$0<\alpha<\pi$]{\includegraphics*[width=.3\linewidth]{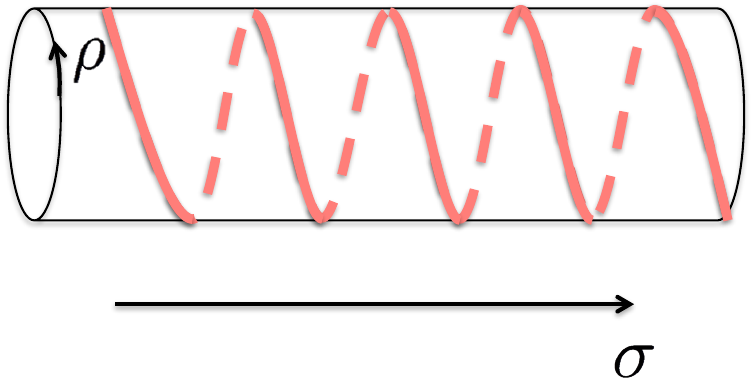}
\label{figurea7c}}
\subfigure[Expansion of the $\sigma$-$\rho$ plane.]{\includegraphics*[width=.6\linewidth]{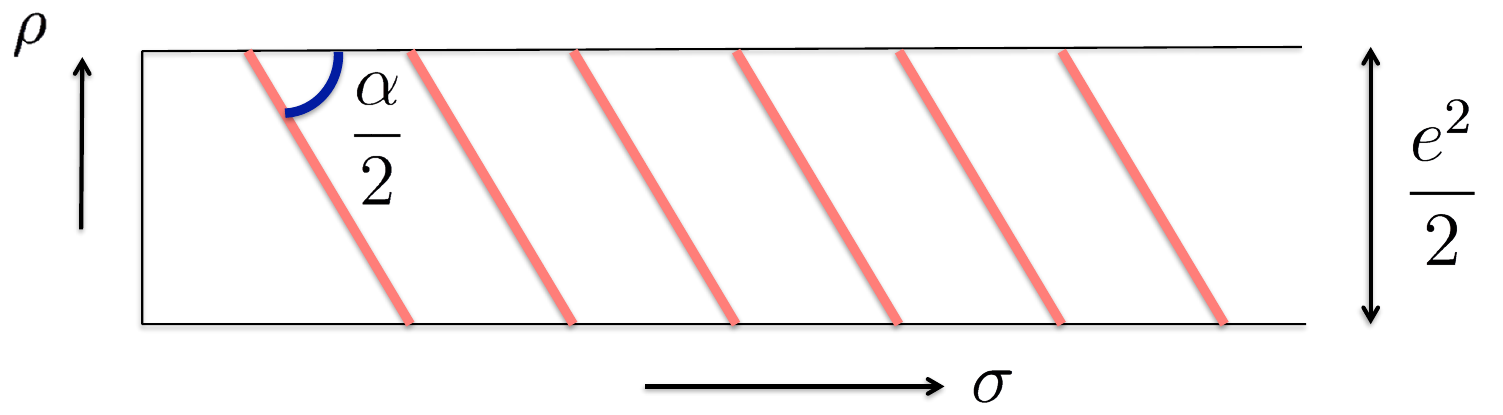}
\label{figurea8d}}
\end{center}
\caption{A-type boundary conditions for
 $\rho$ and $\sigma$. The angle $\alpha$ in a $\sigma$-$\rho$ plane can
 be identified with the phase appearing in the action of $\gamma^{2}$ on $\lambda$.}
\label{figurea71}
\end{figure}

Let us see this boundary condition in the dual picture. (\ref{maxwella}) become
\begin{align}
\partial_{a}\left(
\cos\frac{\alpha}{2}\rho+\sin\frac{\alpha}{2}\sigma
\right)&=0, \nonumber \\
\partial_{2}
\left(
\sin\frac{\alpha}{2}\rho-\cos\frac{\alpha}{2}\sigma
\right)&=0 ,  \qquad\textrm{where} \quad a=0,1.
\end{align}
Since the first condition is interpreted as Dirichlet-type condition,
these conditions can be shown as in Figure \ref{figurea71}.  These branes are actually Lagrangian submanifolds on the cylinder and consistent with the analysis in the section \ref{sec:LG}.

\subsubsection{B-type $\gamma^{2}\epsilon=\epsilon$}
Here we consider two ansatze
\begin{align}
\gamma^2\lambda=+\lambda,
\end{align}
and
\begin{align}
\gamma^2\lambda=-\lambda.
\end{align}
We call these ansatze (BI) and (BII) respectively.

\begin{itemize}
\item[{(BI)}] $\gamma^{2}\lambda=+\lambda$

Noting that $\epsilon\lambda=0$, 
(\ref{purecond1}) becomes
\begin{align}
0=
-\epsilon\sigma^{a}(iF_{2a}\bar{\lambda}+\partial_{a}\sigma\gamma^{2}\lambda)
-\bar{\epsilon}\sigma^{a}(iF_{2a}\lambda-\partial_{a}\sigma\gamma^{2}\lambda).
\end{align}
Therefore we obtain the conditions:
\begin{align}
\label{maxwellb1}
G_{a}&=0, \nonumber \\
\partial_{a}\sigma&=0,  \qquad\textrm{where} \quad a=0,1
\end{align}
The first condition is Neumann-type condition for gauge field $A_{a}$
because it leads to $F_{2a}=0,\ a=0,1$. The second one is
Dirichlet one for scalar $\sigma$. 
This brane is a point on the cylinder and a holomorphic submanifold. This is consistent with the analysis of section \ref{sec:LG}.  
We can also introduce the theta term contribution \eqref{thetaterm1} which expresses the position in the $\rho$ direction.

Let us see this boundary condition in the dual picture. (\ref{maxwellb1}) is expressed as
\begin{align}
\partial_{a}\rho&=0, \nonumber \\
\partial_{a}\sigma&=0,  \qquad\textrm{where} \quad a=0,1.
\end{align}
Now both conditions can be understood as Dirichlet-type conditions for
      $\sigma$ and $\rho$. The configurations are illustrated in Figure
      \ref{figurea8a}.

\begin{figure}[H]
\begin{center}
\includegraphics[width=5cm]{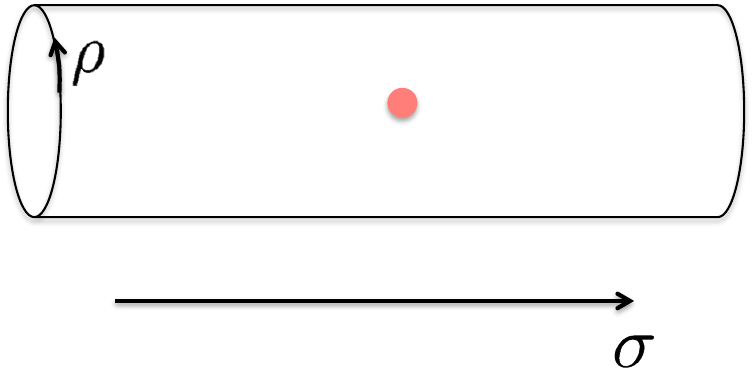}
\caption{BI-type boundary conditions for $\sigma$ and $\rho$. One imposes
 Dirichlet boundary condition on both $\sigma$ and $\rho$. The
 configuration can be expressed as a point in a $\sigma$-$\rho$ plane.}
\label{figurea8a}
\end{center}
\end{figure}

\item[(BII)] $\gamma^{2}\lambda=-\lambda$

Since we have the relation 
$\epsilon\sigma^{a}\lambda=0,\ a=0,1$, 
the condition (\ref{purecond1}) is simplified as
\begin{align}
0=\epsilon\bar{\lambda}
\left(
iF_{01}-\partial_{2}\sigma
\right)
+\bar{\epsilon}\lambda
\left(
iF_{01}
+\partial_{2}\sigma
\right).
\end{align}
This leads to the condition
\begin{align}
\label{maxwellb2}
G_{2}&=0, \nonumber \\
\partial_{2}\sigma&=0.
\end{align}
In this case the first condition is Dirichlet boundary condition for
      gauge field $A_{a},\ a=0,1$ since it requires that $F_{01}=0$. On
      the other hand, the second one is Neumann condition for scalar
      $\sigma$. 
Notice that in this case we have no theta term because $F_{01}=0$.

Let us see this brane in the dual picture. (\ref{maxwellb2}) is rewritten as
\begin{align}
\partial_{2}\rho&=0 \nonumber \\
\partial_{2}\sigma&=0.
\end{align}
In this case both of these are Neumann conditions. 
These are explained in Figure \ref{figurea8b}.  This brane is also a holomorphic submanifold and consistent with the analysis of section \ref{sec:LG}.  This brane extends to the $\rho$ direction and thus consistent with the fact that the boundary does not admit the theta term.

\begin{figure}[H]
\begin{center}
\includegraphics[width=5cm]{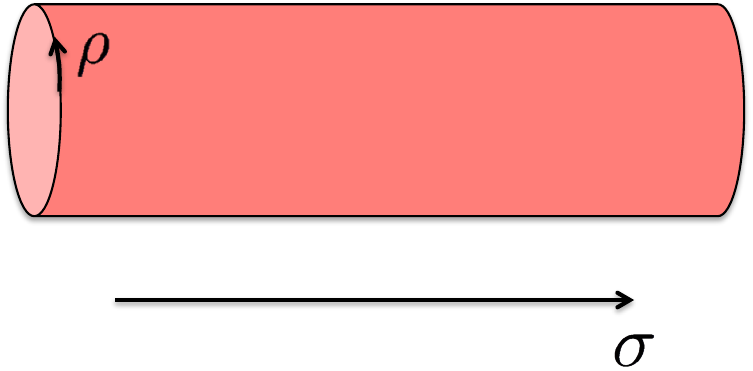}
\caption{BII-type boundary conditions for $\sigma$ and $\rho$. One
 imposes Neumann condition on both $\sigma$ and $\rho$. The configuration
 can be expressed as the entire $\sigma$-$\rho$ plane.}
\label{figurea8b}
\end{center}
\end{figure}

\end{itemize}

%%%%

\section{SQED}
\subsection{Boundary condition of SQED}
Now we want to discuss three-dimensional $\mathcal{N}=2$ supersymmetric electrodynamics (SQED).
In this case, in addition to the vector multiplet $V$, we need to introduce charged chiral
superfields $\Phi_{+}$ and $\Phi_{-}$, whose charges are $+1$ and $-1$
respectively. The Lagrangian of SQED is given by
\begin{align}
\label{qedlagrangian}
\mathcal{L}_{\textrm{QED}}
&=\left[\frac{1}{e^{2}}\Sigma^{2}
-\bar{\Phi}_{+}e^{-2V}\Phi_{+}
-\bar{\Phi}_{-}e^{2V}\Phi_{-}\right]\Bigl|_{-\theta\theta\thetab\thetab} \nonumber \\
&=\frac{1}{e^{2}}
\left(
-\frac{1}{4}F_{\mu\nu}F^{\mu\nu}-i\bar{\lambda}\sigma^{\mu}\partial_{\mu}\lambda
-\frac{1}{2}\partial^{\mu}\sigma\partial_{\mu}\sigma+\frac{1}{2}D^{2}\right)
 \nonumber \\
&\quad-D_{\mu}\phib_{+}D^{\mu}\phi_{+}
-i\psib_{+}\sigma^{\mu}D_{\mu}\psi_{+}
+\bar{F}_{+}F_{+} \nonumber \\
&\quad-i\sigma(\psib_{+}\psi_{+})
-\sqrt{2}i\phi_{+}(\psib_{+}\bar{\lambda})
-\sqrt{2}i\phib_{+}(\psi_{+}\lambda)
-\phib_{+}\phi_{+}D-\phib_{+}\phi_{+}\sigma^{2} \nonumber \\
&\quad-D_{\mu}\phib_{-}D^{\mu}\phi_{-}
-i\psib_{-}\sigma^{\mu}D_{\mu}\psi_{-}
+\bar{F}_{-}F_{-} \nonumber \\
&\quad+i\sigma(\psib_{-}\psi_{-})
+\sqrt{2}i\phi_{-}(\psib_{-}\bar{\lambda})
+\sqrt{2}i\phib_{-}(\psi_{-}\lambda)
+\phib_{-}\phi_{-}D-\phib_{-}\phi_{-}\sigma^{2}.
\end{align}
Here we define the covariant derivatives as
\begin{align}
&D_{\mu}\phi_{\pm}:=\partial_{\mu}\phi_{\pm}\mp i A_{\mu}\phi_{\pm}
 ,\qquad D_{\mu}\psi_{\pm}:=\partial_{\mu}\psi_{\pm} \mp iA_{\mu}\psi_{\pm}  \nonumber \\
&D_{\mu}\phib_{\pm}:=\partial_{\mu}\phib_{\pm}
\pm i A_{\mu}\phib_{\pm} , \qquad 
D_{\mu}\psib_{\pm}:=\partial_{\mu}\psib_{\pm}\pm i A_{\mu}\psib_{\pm}.
\end{align}
%where $q\in \mathbb{Z}$ is implemented as the electric charge.

In Wess-Zumino gauge, the supersymmetric transformation for the chiral
multiplet is given by
\begin{align}
\delta\phi_{\pm}&=\sqrt{2}\epsilon\psi_{\pm} ,\nonumber \\
\delta\psi_{\pm}&=\sqrt{2}i \gamma^{\mu}\epsilon D_{\mu}\phi_{\pm}
+\sqrt{2}\epsilon F_{\pm} \mp \sqrt{2}i\bar{\epsilon}\sigma\phi_{\pm},
 \nonumber \\
\delta F_{\pm}&=\sqrt{2}i\bar{\epsilon}\sigma^{\mu}D_{\mu}\psi_{\pm}
\pm 2i(\bar{\epsilon}\bar{\lambda})\phi_{\pm}\pm \sqrt{2}i(\bar{\epsilon}\psi_{\pm})\sigma.
\end{align}

Then supercurrents for SQED are calculated as 
\begin{align}
{J_{\textrm{QED}}}^{\mu}&=-iF^{\mu\nu} \gamma_{\nu}\bar{\lambda}
+\frac{i}{2}\epsilon^{\rho\sigma\mu}
\bar{\lambda}
F_{\rho\sigma}
+\gamma^{\mu\nu}\bar{\lambda} \partial_{\nu}\sigma
-\bar{\lambda}\partial^{\mu}\sigma \nonumber \\
&\quad-\sqrt{2}D^{\mu}\phib_{+}\psi_{+}
-\phib_{+}\phi_{+}\gamma^{\mu}\bar{\lambda}
+\sqrt{2}D_{\nu}\phib_{+}\gamma^{\mu\nu}\psi_{+}
-\sqrt{2}\sigma\phib_{+}\gamma^{\mu}\psi_{+} \nonumber \\
&\quad-\sqrt{2}D^{\mu}\phib_{-}\psi_{-}
+\phib_{-}\phi_{-}\gamma^{\mu}\bar{\lambda}
+\sqrt{2}D_{\nu}\phib_{-}\gamma^{\mu\nu}\psi_{-}
+\sqrt{2}\sigma\phib_{-}
\gamma^{\mu}\psi_{-} , \nonumber \\ 
{\bar{J}_{\textrm{QED}}}{}^{\mu}
&=iF^{\mu\nu}\gamma_{\nu}\lambda
-\frac{i}{2}\epsilon^{\rho\sigma\mu}
\lambda
F_{\rho\sigma}
+\gamma^{\mu\nu}\lambda\partial_{\nu}\sigma
-\lambda\partial^{\mu}\sigma \nonumber \\
&\quad-\sqrt{2}D^{\mu}\phib_{+}\bar{\psi}_{+}
-\phib_{+}\phi_{+}\gamma^{\mu}\lambda
+\sqrt{2}D_{\nu}\phib_{+}\gamma^{\mu\nu}\bar{\psi}_{+}
-\sqrt{2}\sigma\phib_{+}\gamma^{\mu}\bar{\psi}_{+} \nonumber \\
&\quad-\sqrt{2}D^{\mu}\phi_{-}\bar{\psi}_{-}
+\phib_{-}\phi_{-}\gamma^{\mu}\lambda
+\sqrt{2}D_{\nu}\phi_{-}\gamma^{\mu\nu}\psi_{-}
+\sqrt{2}\sigma\phi_{-}
\gamma^{\mu}\bar{\psi}_{-}.
\end{align}
Thus we obtain the supersymmetric boundary condition for SQED:
\begin{align}
\label{qedcond1}
0&=\epsilon
 J^{2}_{\textrm{QED}}-\bar{\epsilon}\bar{J}^{2}_{\textrm{QED}} \nonumber \\
&=-iF^{2a}(\epsilon\sigma_{a}\bar{\lambda})
+i(\epsilon\bar{\lambda})F_{01}
+(\epsilon C\gamma^{2a}\bar{\lambda})\partial_{a}\sigma
-(\epsilon\bar{\lambda})\partial^{2}\sigma \nonumber \\
&\quad-\sqrt{2}D_{2}\phib_{+}(\epsilon\psi_{+})
-\phib_{+}\phi_{+}(\epsilon\sigma_{2}\bar{\lambda})
+\sqrt{2}D_{a}\phib_{+}(\epsilon C\gamma^{2a}\psi_{+})
-\sqrt{2}\sigma\phib_{+}(\epsilon\sigma_{2}\psi_{+}) \nonumber \\
&\quad-\sqrt{2}D_{2}\phib_{-}(\epsilon\psi_{-})
+\phib_{-}\phi_{-}(\epsilon\sigma_{2}\bar{\lambda})
+\sqrt{2}D_{a}\phib_{-}(\epsilon C\gamma^{2a}\psi_{-})
+\sqrt{2}\sigma\phib_{-}(\epsilon\sigma_{2}\psi_{-})\nonumber \\
&\quad+(c.c.) \qquad\qquad\qquad\qquad\qquad\qquad \textrm{where} \quad a=0,1. 
\end{align}
Here is an example of B-type boundary condition ($\gamma^2\epsilon=\epsilon$):
\begin{align}
\gamma^2\psi_{\pm}=\psi_{\pm},\quad
\gamma^2\lambda=-\lambda,\quad
\phi_{\pm}=0,\quad
\del_{2}\sigma=F_{01}=0.\label{example}
\end{align}

We will not pursue the full classification of the boundary conditions in this paper. Instead let us discuss the mirror symmetry for the above example of the boundary condition.

\subsection{Mirror symmetry and boundary}
\begin{figure}
\begin{center}
\includegraphics[width=12cm]{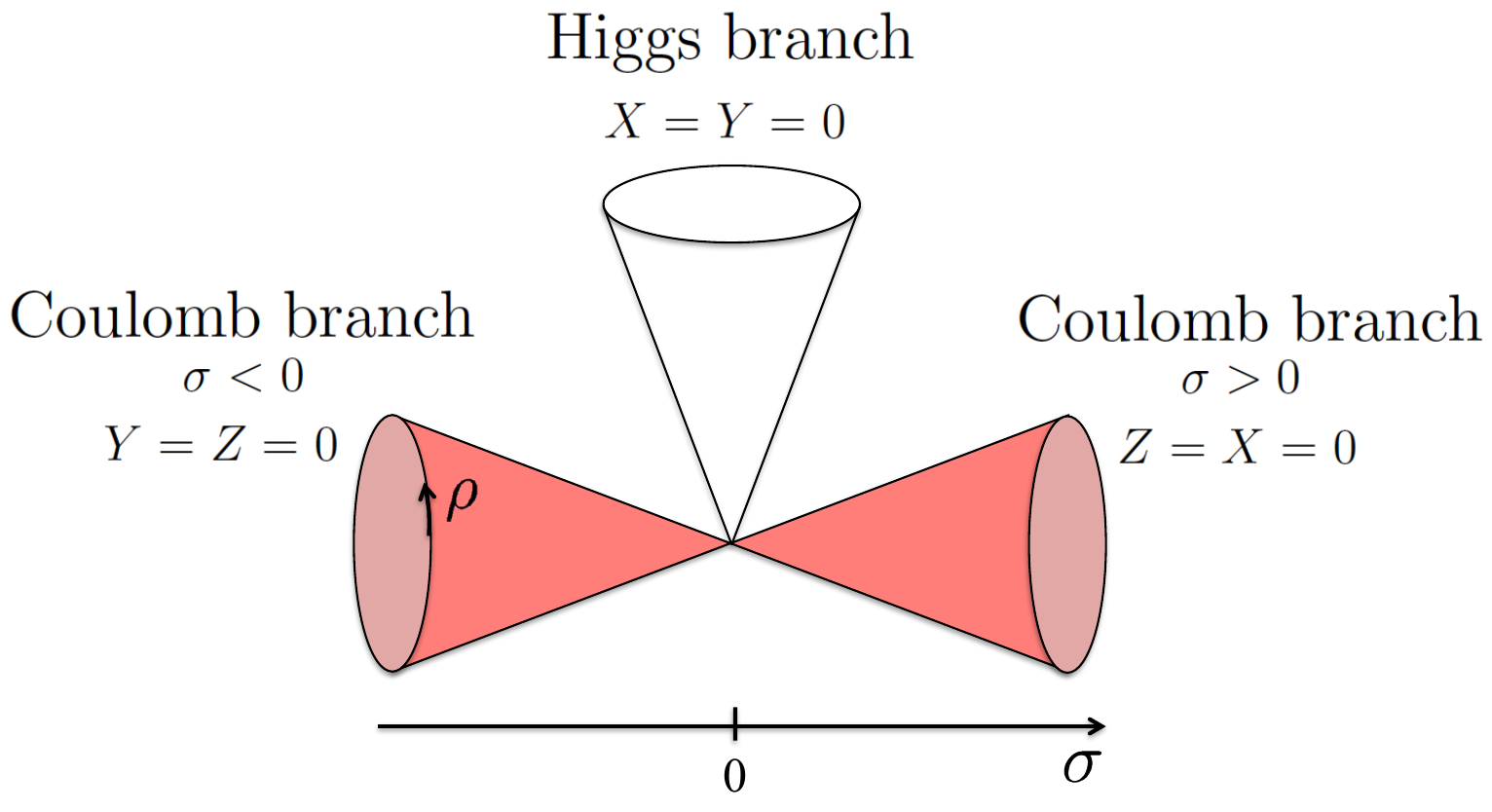}
\end{center}
\caption{The moduli space of SQED and the XYZ model. It contains three branches. They are the Higgs branch, the Coulomb branch with $\sigma>0$ and the Coulomb branch $\sigma<0$ in terms of SQED.  They are $X=Y=0$, $Z=X=0$ and $Y=Z=0$ in terms of the XYZ model.
The example of the brane \eqref{example} is indicated by the red region. It fills the Coulomb branch. It is the same as $Z=0$ region in the moduli space of the XYZ model.}
\label{fig:moduli}
\end{figure}

The SQED considered above is conjectured to be equivalent to ``the XYZ model'' in the low energy limit.  The XYZ model contains three chiral superfields $X,Y,Z$ with the superpotential 
\begin{equation}
W=XYZ.
\end{equation}
This equivalence is called ``mirror symmetry.''
One piece of evidence is that the moduli space of vacua of SQED coincides with that of the XYZ model\cite{Aharony:1997bx,deBoer:1997kr}.  See Figure \ref{fig:moduli}.  

Let us consider how a boundary condition of SQED is mapped to the XYZ model.
Notice that the mirror symmetry in three-dimensions does not exchange an
A-brane and a B-brane.  An A-brane and a B-brane preserve $\Ncal=(1,1)$
SUSY and $\Ncal=(2,0)$ SUSY respectively in 3-dimensions.
Thus an A-brane and a B-brane cannot be mirror to each other.
This is one of the differences between two-dimensions and three-dimensions.
In two-dimensional field theories with boundary, both A-type and B-type preserve 
$\Ncal=2$ on the boundary.  Thus an A-brane can be mirror to a B-brane.

Here we conjecture that the B-type boundary \eqref{example} in SQED corresponds to the B-brane in the XYZ model described by the hypersurface
\begin{align}
Z=0.
\end{align}
This brane is an holomorphic submanifold and the superpotential
$W=0=$(constant) on its world volume. Thus this boundary condition preserves B-type SUSY from the analysis of section \ref{sec:LG}.

An evidence for this correspondence is the location of the brane in the moduli space.  In both sides the brane fills two branches out of three. In the SQED side these two branches are Coulomb branches.  Actually the boundary condition \eqref{example} is Neumann to the Coulomb branch direction spanned by $\sigma$ and the ``dual photon'' $\rho$, while it does not extend to the Higgs branch as seen from $\phi_{\pm}=0$.   In the XYZ model side $Z=0$ brane fills the two branches out of three as seen in Figure \ref{fig:moduli}.

For a B-type boundary a superconformal index is also defined in the same way as \cite{Kim:2009wb,Imamura:2011su,Kapustin:2011jm}.
It will be an interesting future work to calculate the superconformal index with boundary and check whether the SQED result and the XYZ model result agree with each other.  

\section{Conclusion and Discussion}
In this paper, we provided supersymmetric boundary conditions in
three-dimensional $\mathcal{N}=2$ Landau-Ginzburg model and Abelian gauge theories. 
We analyzed the Abelian duality of the boundary conditions between pure
Maxwell theory and chiral matter theory. 
Our result revealed the exact
correspondence in terms of supersymmetric boundary conditions. 
Furthermore we investigated supersymmetric boundary conditions in $\mathcal{N}=2$ SQED, which is supposed to be dual to XYZ model.  We made a conjecture on the mirror dual of an example of B-type boundary condition.

One can expect many possible applications and future directions related to our analysis.
It will be a very interesting future problem to calculate the
superconformal index with a boundary to check the mirror symmetry.  As
discussed in \cite{Gaiotto:2008sa}, supersymmetric boundary conditions
can be identified with BPS domain walls by using the folding trick. A
related calculation of the index in the presence of a domain wall in
four dimensions has been done in \cite{Gang:2012ff}.  It also seems to be an interesting problem to calculate the partition functions of the $\Ncal=2$ theory on other spaces with boundary such as hemispheres and hemiellipsoids by localization \cite{Kapustin:2009kz,Hama:2011ea}.  In particular it should be fruitful to investigate the boundary c-theorem proposed by \cite{Nozaki:2012qd} 

Another interesting future work is to investigate the role of boundaries or domain walls 
in 3d-3d
correspondence\cite{Terashima:2011qi,Terashima:2011xe,Dimofte:2011ju,Dimofte:2011py}.
We expect that 2d SUSY and 4d-non-SUSY versions of AGT relations might be
found by investigating the duality domain walls in three-dimensional
$\Ncal=2$ theories. Recent work in \cite{Gadde:2013wq} is in the same
direction, in which $(2,0)$ supersymmetry, B-type boundary
condition is chosen on the two-dimensional boundary.

Also we would like to understand these results in string theory.  
In \cite{deBoer:1996ck, deBoer:1997ka}, they discuss three-dimensional
       mirror symmetries by using string theory. It is natural to think of
       our problems including boundary in such constructions.

Moreover, in our construction, we have $(2,0)$ and $(1,1)$ supersymmetry on
      the two-dimensional boundary. So far there are few examples of
      mirror symmetries for
      such theories. Our constructions can be useful to explore them.

%%%%%%%%%%%%%%%%%%%%%%%%%%%%%%%%%%%
\subsection*{Acknowledgments}
We would like to thank Tohru Eguchi, Abhijit Gadde, Sergei Gukov,
Kentaro Hori, Tetsuji Kimura, Yu Nakayama, Hirosi Ooguri, Pavel Putrov, Mauricio Romo,
John H. Schwarz, Yuji Tachikawa, and Yutaka Yoshida for discussions and comments.
The work of T.O. was supported in part by JSPS fellowships for Young
Scientists. The work of S.Y. was supported in part by JSPS KAKENHI Grant
No. 22740165.
%%%%%%%%%%%%%%%%%%%%%%%%%%%%%%%%%%

\appendix

\section{Spinors}
In this appendix, we give our notations and useful formulas in
three-dimensional $\mathcal{N}=2$ theories.
We use the metric $\eta_{\mu\nu}=\eta^{\mu\nu}=\mathrm{diag}(-1,1,1)$
and $2\times 2$ $\gamma^{\mu}$ matrices satisfy
\begin{align}
\{\gamma^{\mu},\gamma^{\nu}\}=2\eta^{\mu\nu}.
\end{align}
$\gamma^{0}$ is taken as anti-Hermitian and $\gamma^{1}$ and
$\gamma^{2}$ as Hermitian.

We introduce C matrix $C$, which has the following properties: 
\begin{align}
 &C^{\dag}=C^{-1},\quad C^{T}=-C,\quad (C\gamma^{\mu})^{T}=C\gamma^{\mu}
\end{align}

Two-component spinors $\psi^{\alpha}$ with upper or lower indices transform under $C$:
\begin{align}
& \psi_{\alpha}:=C_{\alpha\beta}\psi^{\beta},\quad
 \psi^{\alpha}=(C^{-1})^{\alpha\beta}\psi_{\beta}.
\end{align}
We use the following summation convention:
\begin{align}
\quad
(\chi\psi):=\chi^{\alpha}\psi_{\alpha}
=\chi^{\alpha}C_{\alpha\beta}\psi^{\beta},
\quad
(\gamma^{\mu}\psi)^{\alpha}
=\gamma^{\mu}{}^{\alpha}{}_{\beta}\psi^{\beta},
&
\quad 
(C\gamma^{\mu}\psi)_{\alpha}=(C\gamma^{\mu})_{\alpha\beta}\psi^{\beta}.
\end{align}

We define $\sigma$-matrices as 
\begin{align}
\quad
\sigma^{\mu}:=C\gamma^{\mu} ,
\end{align}
 and use the summation expression
$\xi\sigma^{\mu}\psi:=\xi^{\alpha}(C\gamma^{\mu})_{\alpha\beta}\psi^{\beta}$.

We define charge conjugation by 
\begin{align}
 \psib^{\alpha}:=(C(\gamma^0)^T)^{\alpha}{}_{\beta}(\psi^{\beta})^{*}.
\end{align}
Here are useful spinor formulas:
\begin{align}
\xi \psi&=\psi\xi, \qquad \xi \sigma^{\mu} \psi=-\psi \sigma^{\mu} \xi ,\nonumber  \\
\psi \sigma^{\mu}\psi&=0, \qquad \psi C\gamma^{\mu\nu}\chi =- \chi C\gamma^{\mu\nu}\psi,
\end{align}

\begin{align}
(\xi \psi)^{\dag}=-\psib\xib, \qquad (\xi \sigma^{\mu}\psi)^{\dag}=\psib \sigma^{\mu}\xib
=-\xib \sigma^{\mu}\psib,
\end{align}

\begin{align}
\theta_{\alpha}\theta_{\beta}=\frac12 C_{\alpha\beta}\theta\theta, \qquad \theta^{\alpha}\theta^{\beta}&=-\frac12
 (C^{-1})^{\alpha\beta}\theta\theta,
\end{align}
\begin{align}
(\theta\psi)(\theta\chi)&=-\frac12(\theta\theta)(\psi\chi) ,\\
(\theta\sigma^{\mu}\chi)(\theta \psi)
&=-\frac12\theta\theta\psi\sigma^{\mu}\chi ,\\
\theta\sigma^{\mu}\thetab \theta \sigma^{\nu}\thetab&=\frac12
 \theta\theta\thetab\thetab \eta^{\mu\nu},
\end{align}

\begin{align}
-\frac12(\chi\lambda)(\psi\xi)
-\frac12(\chi\sigma^{\mu}\lambda)(\psi\sigma_{\mu}\xi)&=
(\chi\xi)(\psi\lambda),
\end{align}

\begin{align}
&C\gamma^{\mu}C^{-1}=-\gamma^{\mu T}
, \qquad C\gamma^{\mu T}C^{-1}=-\gamma^{\mu}.
\end{align}
where $\psi, \xi, \theta, \lambda$ are two-component spinors.

%As usual, Majorana spinors satisfy the condition
%\begin{align}
% \psib^{\alpha}=\psi^{\alpha}.
%\end{align}

\section{Superspace}
We introduce three-dimensional $\Ncal=2$ superspace coordinates 
$(x^{\mu},\theta^{\alpha},\thetab^{\alpha})$, transforming as
$x^{\mu}\rightarrow x^{\mu}-i\epsilon\sigma^{\mu}\thetab
 -i\bar{\epsilon}\sigma^{\mu}\theta$, $\theta\rightarrow \theta+\epsilon$
and $\thetab\rightarrow \thetab+\bar{\epsilon}$
 under the supersymmetry transformations.
% with parameter $\epsilon$ and
%$\bar{\epsilon}$, 
We also define the following supersymmetric derivatives:
\begin{align}
 Q_{\alpha}&:=\deldel{}{\theta^{\alpha}}-i(\sigma^{\mu}\thetab)_{\alpha}\del_{\mu},\\
 \Qb_{\alpha}&:=-\deldel{}{\thetab^{\alpha}}+i(\sigma^{\mu}\theta)_{\alpha}\del_{\mu},\\
 D_{\alpha}&:=\deldel{}{\theta^{\alpha}}+i(\sigma^{\mu}\thetab)_{\alpha}\del_{\mu},\\
 \Db_{\alpha}&:=-\deldel{}{\thetab^{\alpha}}-i(\sigma^{\mu}\theta)_{\alpha}\del_{\mu}.
\end{align}
They have the anticommutation relations
\begin{align}
 \{Q_{\alpha},\Qb_{\beta}\}=2i\sigma^{\mu}_{\alpha\beta}\del_{\mu}, 
\quad \{D_{\alpha},\Db_{\beta}\}=-2i\sigma^{\mu}_{\alpha\beta}\del_{\mu},
\end{align}
with all the other anticommutators vanishing. 
The supersymmetry transformation of a
superfield $\Phi(x,\theta,\thetab)$ is expressed as
\begin{align}
 \delta \Phi(x,\theta,\thetab)=(\epsilon Q-\epsilonb \Qb)\Phi.
\end{align}

\section{Superfield}
\subsection{Chiral superfield}
Chiral superfield $\Phi(x,\theta,\thetab)$ is defined as
\begin{align}
 \Db_{\alpha}\Phi=0.
\end{align}
Using $y^{\mu}:=x^{\mu}+i\theta \sigma^{\mu} \thetab$, we obtain the
component field representations:
\begin{align}
 \Phi&=\Phi(y,\theta) \nonumber \\
&=\phi(y)+\sqrt{2}\theta\psi(y)+\theta\theta F(y) \nonumber \\
&\label{chiralcomponent}
=\phi(x)+i\theta\sigma^{\mu}\thetab \del_{\mu}\phi(x)-\frac14 \theta\theta \thetab\thetab \del^2 \phi(x)
+\sqrt{2}\theta\psi(x)+\frac{i}{\sqrt{2}}(\theta\theta) ( \thetab\sigma^{\mu}\del_{\mu}\psi(x))+\theta\theta F(x).
\end{align}
Antichiral superfield $\bar{\Phi}(x,\theta, \thetab)$ with the constraint $D_{\alpha}\bar{\Phi}=0$
 can be obtained from (\ref{chiralcomponent})
by conjugation:
\begin{align}
\bar{\Phi}
=\phib(x)-i\theta\sigma^{\mu}\thetab\partial_{\mu}\phib(x)
-\frac14 \theta\theta\thetab\thetab\partial^{2}\phib(x)
-\sqrt{2}\thetab\psib(x)
-\frac{i}{\sqrt{2}}(\thetab\thetab)(\theta\sigma^{\mu}\partial_{\mu}\psib(x))
-\thetab\thetab \bar{F}(x).
\end{align}

\subsection{Vector superfield}
Vector superfields satisfy the relation
\begin{align}
V=\bar{V}.
\end{align}
Choosing Wess-Zumino gauge we obtain the simple
expression:
\begin{align}
&V=-\theta\sigma^{\mu}\bar{\theta}A_{\mu}
+i\theta\bar{\theta}\sigma
-i\theta\theta\thetab\bar{\lambda}
+i\thetab\thetab\theta\lambda+\frac12\theta\theta\thetab\thetab D(x).
\end{align}

We can express field strength as a linear multiplet:
\begin{align}
\Sigma:=-\frac{i}{2}\bar{D}DV.
\end{align}
In the component description, it is written as
\begin{align}
\Sigma&=\sigma+\theta\bar{\lambda}-\lambda\thetab-i(\thetab\theta)D
+\frac{1}{2}(\thetab C\gamma^{\mu\nu}\theta)F_{\mu\nu} \nonumber \\
&-\frac{i}{2}\theta\theta (\thetab\sigma^{\mu}\partial_{\mu}\bar{\lambda})
+\frac{i}{2}\thetab\thetab(\theta\sigma^{\mu}\partial_{\mu}\lambda)+
\frac{1}{4}\theta\theta\thetab\thetab\partial^{\mu}\partial_{\mu}\sigma.
\end{align}

\bibliographystyle{utphys}
\bibliography{ref}

\end{document}